\documentclass[graybox]{svmult}

\usepackage{mathptmx}       
\usepackage{helvet}         
\usepackage{courier}        
\usepackage{type1cm}        
\usepackage{makeidx}         
\usepackage{graphicx}        
\usepackage{multicol}        
\usepackage[bottom]{footmisc}

\makeindex             

\def\msol{M_\odot}

\def\tmdot{\tau_{\rm \dot M}}
\def\tkh{\tau_{\rm KH}}

\def\msun{M_\odot}

\def\te{T_{\mathrm{eff}}}

\def\be{\begin{equation}}
\def\ee{\end{equation}}

\def\simgr{\,\hbox{\hbox{$ > $}\kern -0.8em \lower 1.0ex\hbox{$\sim$}}\,}
\def\simle{\,\hbox{\hbox{$ < $}\kern -0.8em \lower 1.0ex\hbox{$\sim$}}\,}

\begin{document}

\title*{Latest news on the Physics of Brown dwarfs}
\author{Isabelle Baraffe}
\institute{Isabelle Baraffe \at University of Exeter, Physics and Astronomy, EX4 4QL Exeter UK, \email{i.baraffe@exeter.ac.uk}}
%
%
\maketitle


\abstract{The physics of brown dwarfs has continuously improved since the discovery of these astrophysical bodies. The first important developments were devoted to the description of their mechanical structure, with the derivation of an appropriate equation of state, and the modelling of their atmosphere characterised by strong molecular absorption.  New challenges are arising with progress in observational techniques which provide data of  unprecedented accuracy. The goal of this chapter is to describe some of the current  challenges for the theory of brown dwarfs. Those challenges concerns  atmospheric dust and cloud, non-equilibrium atmospheric chemistry, the effect of rotation and magnetic fields on internal structure and the very early phases of evolution characterised by accretion processes. The field remains lively as more and more high quality observational data become available and because of increasing discoveries of exoplanets. Indeed, many physical properties of giant exoplanets can be described by the same theory as brown dwarfs, as described in this chapter. }

\section{Introduction: the first theoretical challenges}
\label{intro}

 The history of brown dwarfs is particularly interesting, since their existence was theoretically predicted  in 1963 before they were discovered in 1995. Coincidently at the same time, \cite{kumar63} and \cite{hayashi63} demonstrated that objects less massive than $\sim$ 0.08-0.09 $\msun$ should have their internal structure affected by electron-degeneracy as they contract gravitationally after their formation. Those objects should be unable to release enough energy from hydrogen nuclear burning, since their central temperatures begin to decrease after degeneracy proceeds. They should never go through the usual stellar evolution phases, living as "failed" stars. The theoretical calculations of the two pioneer papers in 1963 marked the birth of brown dwarfs fifty years ago.  Following these predictions, the motivation was there for more theoretical calculations and evolutionary models for brown dwarfs \cite{grossman70, straka71, rappaport84}, followed by many others. Models based on up-dated micro-physics now predict that the mass limit for the onset of electron-degeneracy is close to 0.07 $\msun$ for objects with a solar metal content (i.e the so-called metallicity) and increases with decreasing metallicity \cite{cb00}.

With the discovery of the first genuine brown dwarfs about 20 years ago, (see the chapters by  R. Rebolo, G. Basri, and by B. Oppenheimer in this volume), the theory and modelling of brown dwarfs have constantly developed and improved. The first major theoretical challenge was the development of an equation of state (EOS hereafter) appropriate for the description of the dense and relatively cool interior of brown dwarfs, accounting for the effects of electron partial degeneracy and interaction between particles, namely molecules, atoms, ions and electrons 
(see  \cite{cb00} and references therein).  The other major challenge was the development of "cool" atmosphere models characterised by strong molecular absorptions and, for the coolest objects, by condensation processes (see \cite{allard97} and references therein).  
 
As brown dwarf  observations deliver data with increasing accuracy and level
of details, modellers need to increase the sophistication of the theory, requiring new physical ingredients and processes to face the reality of observations. This paper, far from being a comprehensive review of  the physics of brown dwarfs, will  focus on some recent developments and novelties which I find exciting as they reflect a thriving and innovative  field.  This choice will hopefully provide guidance and motivate future generation of students and young researcher to work on the theory of brown dwarfs. The paper is organised as follows. I will first discuss current developments in atmosphere and inner structure modelling respectively. I will then discuss young brown dwarfs and current ideas regarding their
very early evolution. I will conclude by providing some of the important topics to develop in the coming years in order to keep the field of brown dwarf active and "young".

\section{Recent advancement in the modelling}
\label{advancement}

\subsection{Atmosphere models}
\label{atmos}

After the discovery of the first brown dwarfs in the 1990s, Teide 1, PPl15 and Gliese~229B \cite{rebolo95, nakajima95, oppenheimer95, basri96}  huge advancement in the field was provided by  new generations of  atmosphere models accounting for  improved treatment of molecular opacities. The atmospheres of objects with effective temperatures cooler than $\te \simle$ 4000K (e.g low mass stars, brown dwarfs) are characterised by the formation of molecules (H$_2$, TiO, VO, H$_2$O, CO, FeH, CaH, etc...).  The strongly wavelength-dependent absorption coefficients resulting from molecular line transitions, along with Rayleigh scattering resulting from light scattering from molecules and characterised by a wavelength dependence $\propto \nu^4$, yield a strong departure of the spectra of cool objects from a black-body energy distribution
(see e.g \cite{allard97}). Most brown dwarfs are characterised by effective temperatures $\te \simle 2000K$,  except for very young massive brown dwarfs which can be hot enough to exceed this temperature. Below  $\te \simle 2000K$, some molecules condense into liquid and solid phases, forming grains and depleting the atmospheric gas phase of a number of molecules (e.g TiO which will be sequestered in more complex compounds such as perovskite CaTiO$_3$).  Along with the drastic modification of the chemical composition of the atmosphere, atmospheric heating resulting from the large grain-opacity (the so-called green house or backwarming" effect)  strongly impacts the  spectral energy distribution of "dusty" brown dwarfs (see more details in the chapter by M. Cushing in this volume).
Cool atmosphere models now include more and more complete molecular linelists and better methods to handle many millions of transitions thanks to growing computer facilities (see \cite{allard97}). 
   Continuous efforts are devoted to the calculation of more accurate molecular opacities, both in the optical, with main absorbers like TiO and VO, and in the near-infrared (e.g H$_2$O, CO, CH$_4$), providing more and more realistic atmosphere models. Those progress are made possible with the development of comprehensive molecular linelists, combining  ab initio quantum mechanical treatment and experimental data, which account for the multitude of rotational, vibrational and electronic transitions (see e.g \cite{tennyson12}). 
About a decade ago, when brown dwarfs were in their 40s,  a breakthrough in the observations of brown dwarfs occurred with the infrared sky surveys DENIS \cite{epchtein97}, SDSS \cite{york00} and 2MASS \cite{skrutskie06} which  provided a wealth of  "dusty" L-dwarfs  and  "methane" T-dwarfs. Those projects provide key information on the M/L transition, from "clean" to "dusty" atmospheres, and on the L/T transition, characterised by complete or incomplete clearing of dust. Many new questions on dust properties and on its evolution with decreasing effective temperatures were raised and are still open.
Similar breakthrough is now occurring with  the wide-field survey WISE \cite{wright10}  which pushes the frontiers of cold, isolated object detection further away.  The doors of the Y-dwarfs realm are now ajar, revealing the promised land of water and ammonia condensation processes.  The coldest isolated object discovered at the time of this writing, the Y dwarf WISE 1828+2650 \cite{cushing11, beichman13}, could have an effective temperature as low as  $\te \sim$ 250K, bringing brown dwarfs  closer and closer to the world of giant planets\footnote{Jupiter has an effective temperature  $\te \sim$ 120k \cite{hubbard99}.}.   

\subsubsection{Coud models and chemistry}
\label{cloudmodels}

Interpretation of observations of L, T and Y dwarfs called for new challenges in atmosphere modelling, specifically regarding cloud models and chemistry calculations. Different approaches for dust models with varying complexity exist in the brown dwarf community. They are described in detail  in \cite{helling08} (see also \cite{marley13}) and we will briefly summarise below the essential ingredients of each approach.  

A simple approach of dust modelling extensively developed and used by Tsuji and collaborators \cite{tsuji02, tsuji04} assumes that dust forms in a restricted region defined by two temperatures. First by a condensation temperature, below which species condensation is allowed, and which will characterise the base of a dust layer. And second by a  critical temperature which is a free parameter and defines the top of the dust layer. In these models, the dust particles are assumed to have constant size. A similar approach has been adopted recently in the models of \cite{barman11a}, but using pressure instead of temperature to define the cloud localisation and thickness. It was used to study the near-infrared spectra of the planetary mass objects HR8799b \cite{barman11a} and 2MASS1207b \cite{ barman11b}. Though lacking  the complexity of dust physics, those approaches offer the valuable advantage of allowing parameter exploration and of qualitatively understanding  the effect of dust  \cite{marley13}.  

More sophisticated approaches to treat dust have been developed, like the models of \cite{ackerman01} which allow for vertical variations of particle number densities and sizes. These models account for the  downward transport of particles due to sedimentation through an  efficiency parameter $f_{\rm sed}$, and for the upward mixing of vapor and condensates via a parameter $K_{\rm zz}$ characterising the vertical eddy diffusion coefficient.  In the models of \cite{allard01}, yielding the widely used COND and DUSTY models, condensation,  coagulation  and sedimentation effects are treated within the diffusion approximation following \cite{rossow78}. These models describe the limiting effects of cloud formation. An additional sophistication to the models of \cite{allard01}, yielding the so-called BT-Settl models, is provided by an improved description of atmospheric mixing processes  (e.g overshooting and convective mixing) based on 2D radiation hydrodynamics simulations \cite{freytag10}. The results of those simulations are used to prescribe a  diffusion coefficient characterising the mixing processes. The most detailed cloud model developed to date is based on the work by \cite{helling06, helling08}. It includes microphysics of grain growth and  relaxes the usual assumption of phase equilibrium between gas and cloud particles. 

\begin{figure}[h]
\sidecaption
\includegraphics[scale=.50]{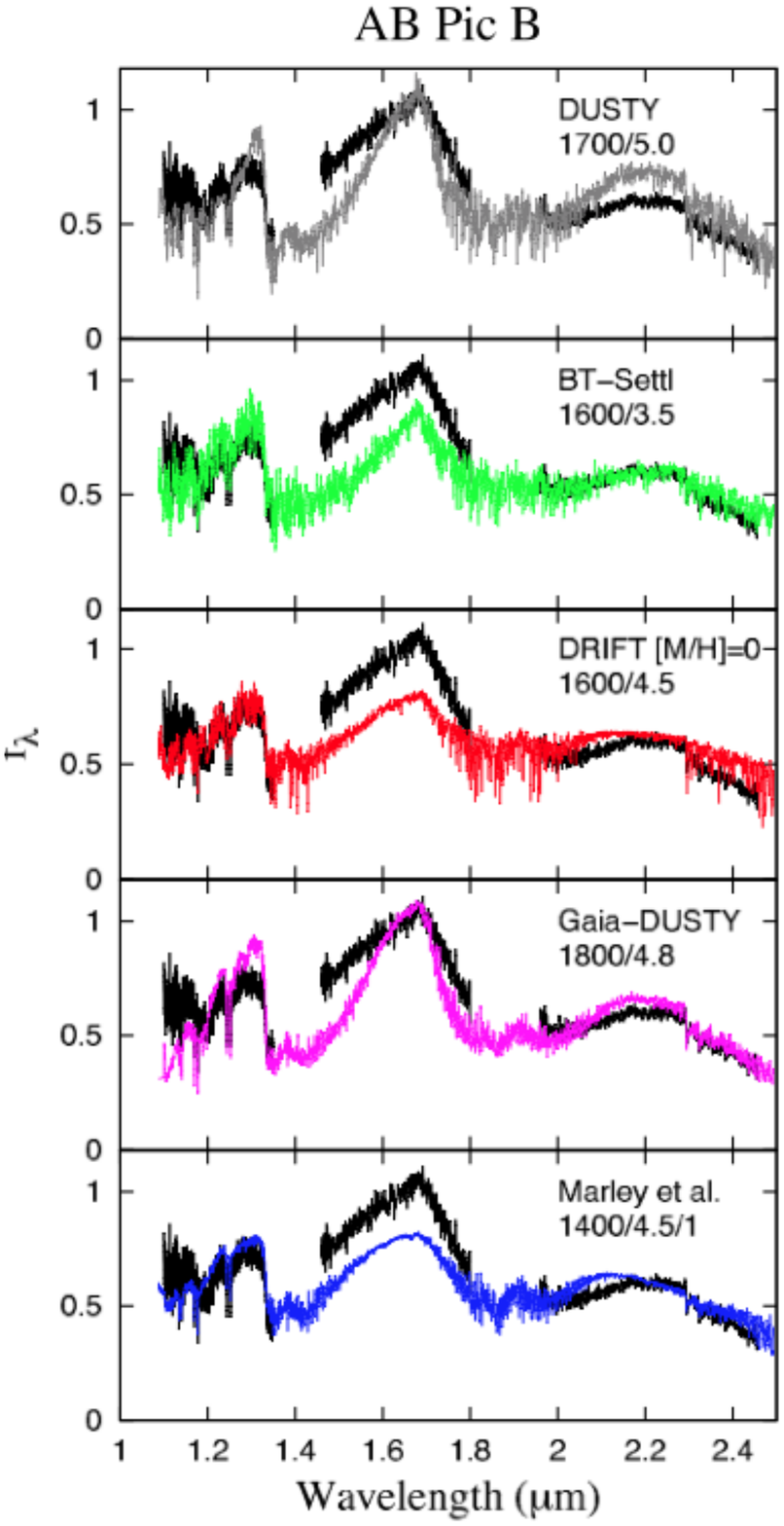}
%
%
\caption{Comparison of the observed spectrum of AB PIC B (black line) with various sets of models.  The effective temperature and the surface gravity $\log g$ inferred by such comparison are indicated in each panel for each model used. One note the large differences between the  effective temperatures and gravities inferred from different models. One also note that none of the models can correctly reproduce the observed spectrum. Figure adopted from  \cite{patience12}.}
\label{plot_patience}       
\end{figure}

All these different approaches have been extensively compared one to another in \cite{helling08}, showing the limit and complexity of dust treatment in cool atmospheres. Interestingly enough, these various models have also been systematically compared to the same observations of young isolated field objects in \cite{patience12} . Figure \ref{plot_patience} shows such comparison for the young early L-dwarf AB PIC B. This work clearly highlights the important effects of dust treatment on synthetic spectra.  The figure shows that none of the models can reproduce the observed spectrum. Additionally, the effective temperatures and surface gravities $g$ (in log)  inferred by the comparison between model and observation are indicated in the figure panels for each model used. The predicted effective temperature varies between 1400K and 1800K and the logarithm of the gravity varies between 3.5 and 5, depending on the model used. This highlights the huge uncertainty of physical properties ($\te$, $\log g$) inferred from a comparison between observed and synthetic spectra for dusty  objects.  It also shows the remaining uncertainties of current models because of their difficulty, no matter how sophisticated the approach is, to reproduce  the observed spectra of objects where dust is expected to form.    
Dust models for brown dwarf (and exoplanet) atmospheres are thus still in their infancy, despite fifty years of theoretical and observational studies in this field. There is still a long way to go to understand and correctly describe the effect of dust  in those objects.

\begin{figure}[h]
\sidecaption
\includegraphics[scale=.40]{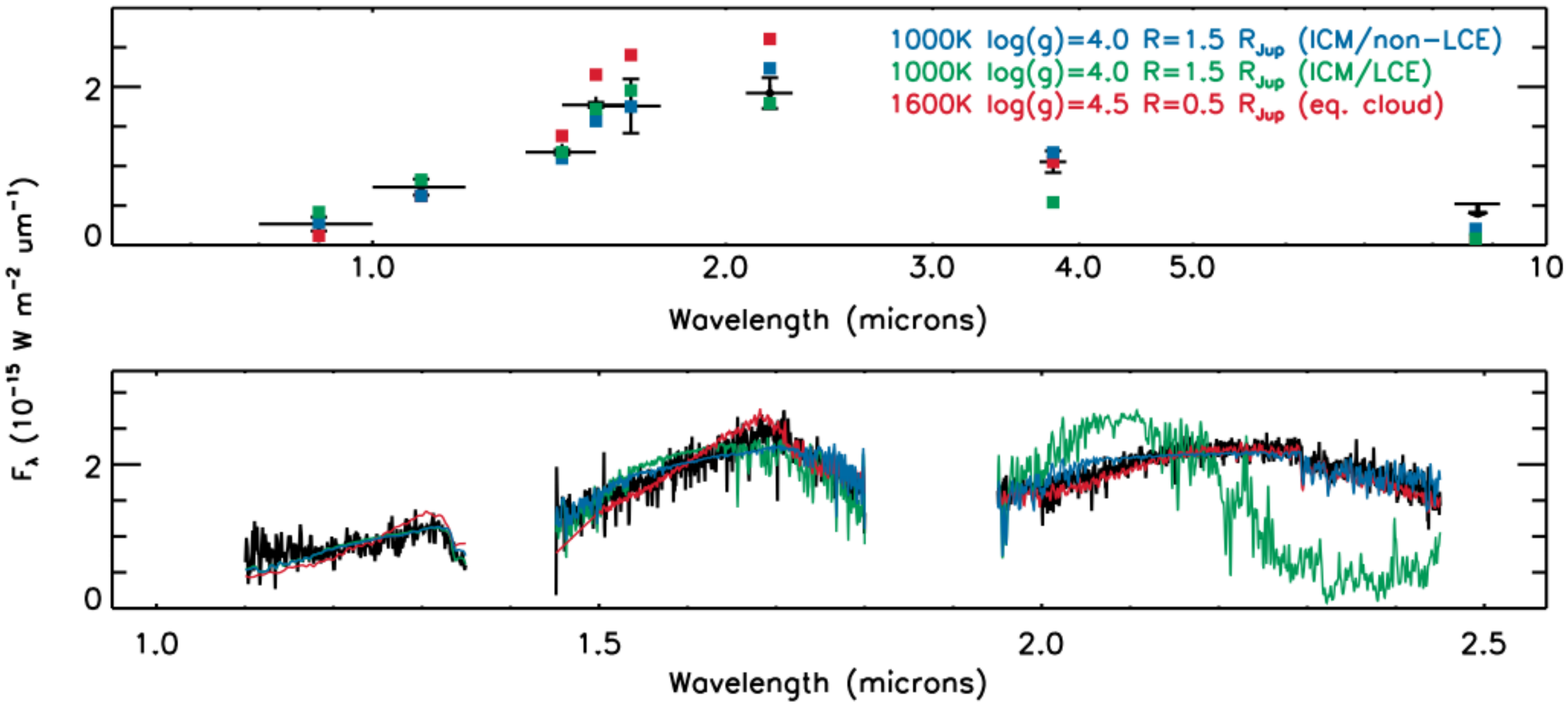}
%
%
\caption{Comparison between observations of the planetary mass object 2M1207b (in black) and models  assuming equilibrium chemistry (green, labelled LCE) and non-equilibrium chemistry (blue, labelled non-LCE) (top: photometry; bottom: spectroscopy). 
 Both models are based on the work of \cite{barman11b} including the cloud model described in \cite{barman11a} (see also Sect.  2.1.1).   These comparisons show the important effect of non-equilibrium chemistry on photometry and spectrum of cool objects (the inferred effective temperature in the present case is $\te \sim 1000K$). 
For comparison, predictions from a model based on a equilibrium cloud model from \cite{allard01} are indicated in red.  Note the large discrepancy between $\te$ inferred from the latter model ($\te \sim 1600$K) and the models of \cite{barman11b}.
Figure from \cite{barman11b} and reproduced by permission of the AAS.}
\label{plot_barman}       
\end{figure}

 Observations of cool brown dwarfs show that, in addition to cloud treatment, non-equilibrium chemistry is also an important process to account for in the models. Departure from equilibrium chemistry in cool atmospheres may be due to various processes, such as irradiation effects in close-in exoplanets, or mixing processes.  The latter may take place in brown dwarf atmosphere if the timescale for vertical mixing is shorter than a chemical reaction timescale. It is expected to occur for the reactions converting  carbon monoxide (CO) to methane (CH$_4$) and nitrogen (N$_2$) to ammonia (NH$_3$). The expected observable signatures of this process  are overabundance of CO and depletion of NH$_3$. Its existence has long been established in jovian planets in our solar system (see e.g \cite{barshay78}). For brown dwarfs, predictions of non-equilibrium chemistry was suggested \cite{fegley96} shortly after the discovery of Gliese~229B  \cite{nakajima95, oppenheimer95} and confirmed with the detection of CO in its infrared spectrum  \cite{noll97}. Evidences that this process is at work in cool atmosphere accumulated since then. 
This was again recently  illustrated by the study of \cite{barman11b} showing that a combination of clouds, given  some prescribed thickness, and non-equilibrium chemistry of CO/CH$_4$ could both reproduce the photometric and spectroscopic observations of the young planetary mass object 2M1207b, as illustrated in Fig. \ref{plot_barman}.
The idea of departure from non-equilibrium chemistry in brown dwarfs is not recent, 
but it now  seems to be a necessary ingredient to include in the models in order to reproduce satisfactorily observed spectra of L and T dwarfs. Similar conclusions was reached in the study of \cite{burningham11} in order to reproduce the infrared spectrum of the T8.5 dwarf Ross 458C with an effective temperature of $\te \sim 700$K. 

\begin{figure}[h]
\sidecaption
\includegraphics[scale=.40]{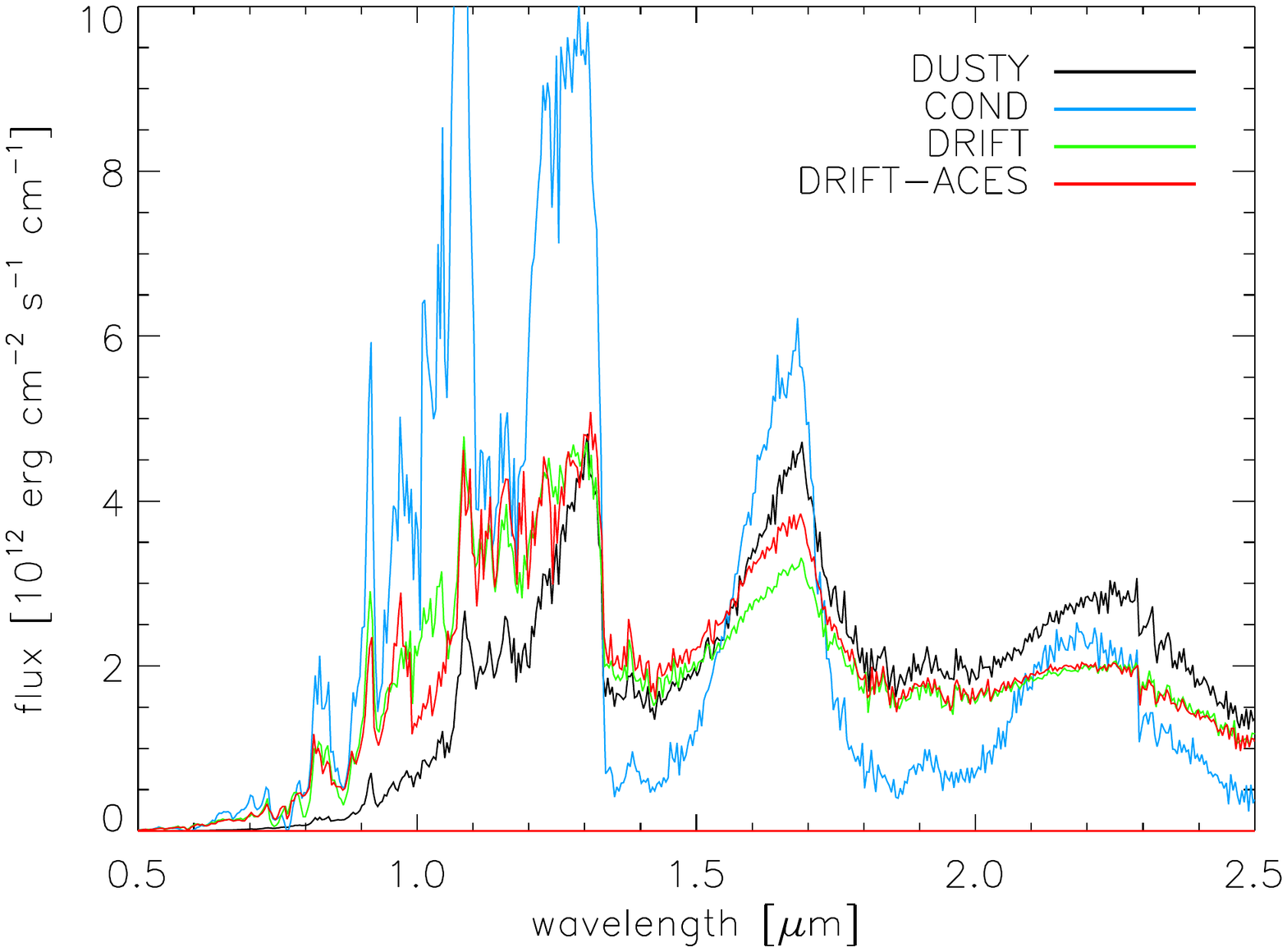}
%
%
\caption{Comparison of synthetic spectra based on different treatments of dust (DUSTY, COND \cite{allard01} and DRIFT \cite{witte09}) and gas-phase chemistry (DRIFT versus DRIFT-ACES \cite{witte11}). Courtesy C. Helling.}
\label{driftaces}       
\end{figure}

Finally, improvements are still made on the treatment of gas-phase chemistry, as illustrated in Fig.  \ref{driftaces} which compares the effect of different EOSs for the gas phase on synthetic spectra (DRIFT versus DRIFT-ACES \cite{witte09, witte11}). The main difference between  DRIFT (green curve  in Fig.  \ref{driftaces}) and DRIFT-ACES (red curve in Fig. \ref{driftaces} ) stems from the treatment of the
gas-phase composition in chemical equilibrium. The ACES improvement consists of a coherent
fit of gas-phase material data for lower temperatures. Hence, the differences observed in this figure between DRIFT and DRIFT-ACES stem alone from the treatment of the gas-phase chemistry (C. Helling priv. com.).

\begin{figure}[!h]
\sidecaption
\includegraphics[scale=.50]{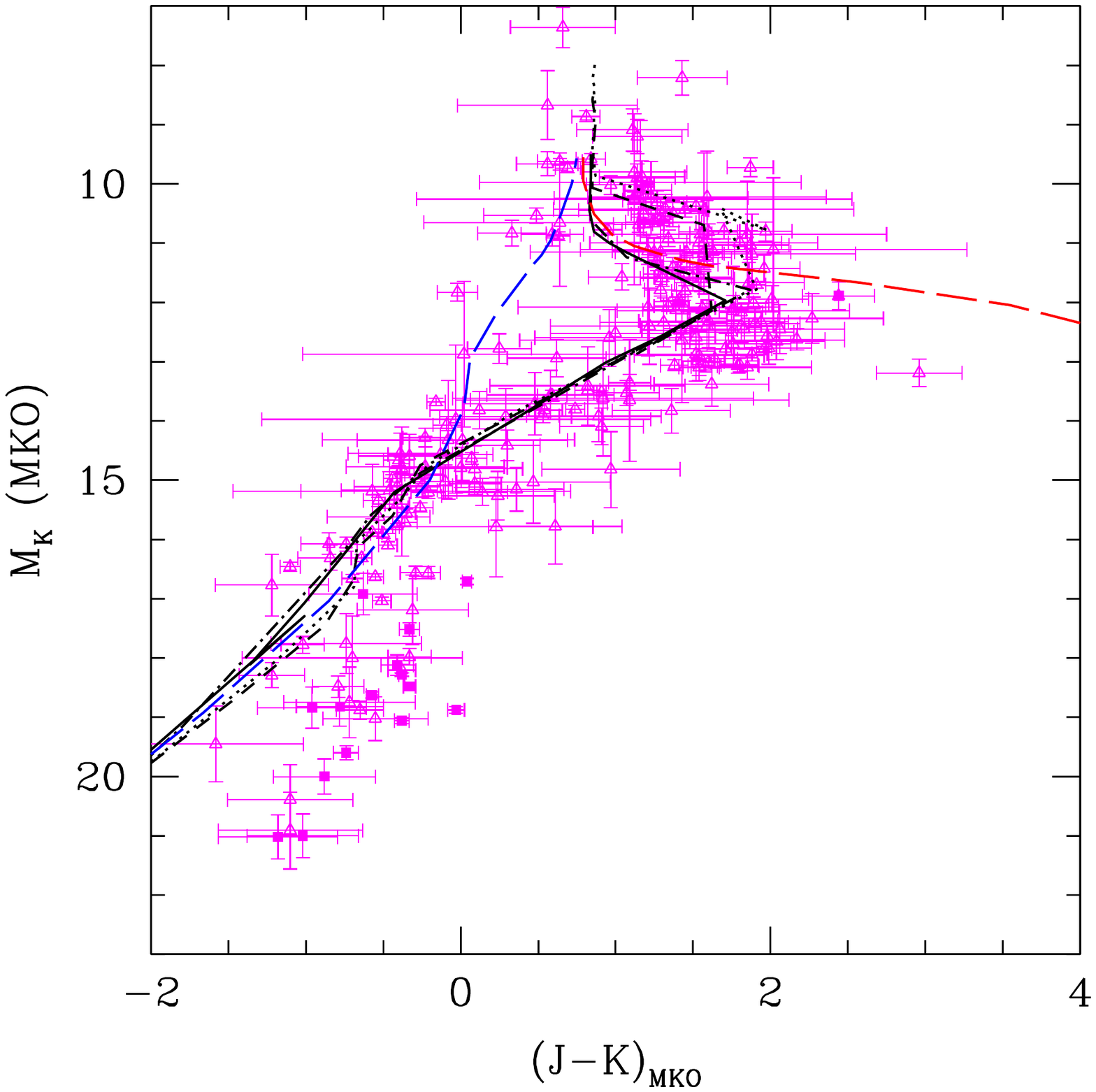}
%
%
\caption{Comparison between models and M-L-T dwarfs data in a (J-K)-M$_{\rm K}$  diagram. The black models are based on the BT-settl atmospheres (\cite{allard12, baraffe13}). Isochrones for 0.05, 0.1, 1 and 5 Gyr are displayed. The Dusty (red) and COND (blue) models at an age of 1 Gyr are also shown for comparison. The data are from \cite{faherty12, dupuy12}.}
\label{jkk}       
\end{figure}

\subsubsection{The L/T transition in the colour-magnitude diagram}
\label{cmd}

The description of the L/T transition in a colour-magnitude diagram is a real challenge for theory.  This transition is physically characterised by the clearing of dust due to  sedimentation of condensed species below the photosphere and the formation of CH$_4$,
which is the dominant equilibrium form of carbon below a local temperature of $\simle 1600$K while CO dominates at higher temperatures. The heating of the atmosphere due to  the backwarming effect of dust causes  infrared colors to become very red, as seen in Fig. \ref{jkk}, whereas the formation of CH$_4$,  which strongly absorbs near 1.6 $\mu$m yields the characteristic change  to bluer (J-H) and (J-K) colors (see the chapter by M. Cushing, his Sect.  7).
In addition to this well-known change in colours from the red to the blue characteristic of the L/T transition in the near-infrared (see Fig. \ref{jkk}),  another major difficulty for models  is to describe the scatter of colours for a given magnitude. Several ideas have been suggested to explain this observed property, namely the effect of a single second parameter like gravity \cite{burrows06} or a mixture of effects, namely metallicity, cloud parameters, ages and unresolved binaries \cite{saumon08}. In the later models, the sedimentation of condensates is characterised by the parameter $f_{\rm sed}$ as introduced by \cite{ackerman01} (cf Sect.  \ref{cloudmodels}). Small values of $f_{\rm sed}$ indicate that particles grow more slowly and thus the condensate load is larger and clouds thicker. Large values of $f_{\rm sed}$ correspond to rapid particle growth, with large condensates quickly falling out of the atmosphere and yielding optically and physically thin clouds. A value of $f_{\rm sed} \sim 2$ is required to describe the L/T transition (see \cite{saumon08}). In comparison, the BT-Settl models  are well reproducing the L/T transition.  The latter models have the advantage of having  no free parameters to describe the sedimentation/mixing processes, which
instead rely on hydrodynamical simulations (see Sect.  \ref{cloudmodels}). But the drawback of using a more physical description of sedimentation in the BT-Settl models is the failure to reproduce the observed scatter of colours. Inspection of Fig. \ref{jkk} also shows a disagreement between the BT-Settl models and observations for  objects fainter than an infrared brightness in the K-band of $M_{\rm K} \sim 17$, with the infrared color (J-K) remaining almost constant as a function of $M_{\rm K}$, whereas  models predict significantly bluer (J-K) colours. This discrepancy is inherent to all current models where the silicate and iron clouds which shape the L dwarf spectra are predicted to dissipate at the L/T transition. 
Recently, \cite{morley12} suggested the formation of other condensates in the coolest T and Y dwarf atmospheres. This solution is attractive since it avoids requiring the reemergence of the silicate clouds, as suggested by \cite{burgasser10} to explain the spectrum of the T8 brown dwarf Ross 458C. 
Those new condensates include sulfide clouds and other species such as Cr and KCl. With a variation of the $f_{\rm sed}$ parameter between $\sim$ 3-5, models including those new clouds provide a better agreement with observed colours for the coolest T-Y dwarfs \cite{morley12}.  The weakness of these models stems from the larger  $f_{\rm sed}$ parameter than the one required to reproduce the L/T transition. Such a change in sedimentation efficiency now requires a physical explanation in order  to provide a satisfactory solution to the discrepancy between models and observations of the coolest dwarfs.  This is still an open issue which 
generates  many efforts  from modellers to understand this puzzle.

\subsection{Inner structure models}
\label{inner}

\subsubsection{Mechanical structure}

 Brown dwarf interiors are  composed of a mixture of hydrogen (H) and helium (He), with traces of metals (i.e elements heavier than He). The thermodynamical properties of such mixture are described by an EOS which provide the relationship between state variables like pressure, temperature and volume (or density). The major fraction of the life of a brown dwarf is characterised by the release of its gravitational and internal energy resulting from a hot initial state after its formation. Only during the very first stage of their evolution, brown dwarfs with masses greater than $\sim$ 0.012 $\msun$, the so-called deuterium burning minimum mass, can  produce nuclear energy due to deuterium fusion in their center. This nuclear phase is however very short, lasting less than $\sim$ 20 Myr. 
The main process which transports energy from the deep hot layers to the surface where it is radiated away, producing the  observable luminosity, is due to convection. The interior of a brown dwarfs is so dense and optically thick that energy transport by radiation is totally negligible.
The mechanical structure (which defines the mass-radius relationship) and internal thermal profile of a brown dwarf are thus determined by the EOS of its chemical constituent. Indeed, because interiors of brown dwarfs are essentially fully convective and radiation is negligible, the resulting thermal profile is quasi-adiabatic, with the adiabatic gradient  fixed by the EOS \cite{cb00}. The most widely used EOS to describe the thermodynamic properties of brown dwarf interiors is the semi-analytical Saumon-Chabrier-VanHorn EOS \cite{sc95} for 
H/He mixtures. Continuous progress is achieved in this field thanks to the growth of computational performances which allow  EOS calculation from first principle quantum mechanical calculations. Progress is also achieved thanks to high-pressure experiments  which allow to probe EOS in a complex regime where pressure-dissociation and -ionisation occur, which characterise the interiors of both brown dwarfs and giant planets. The internal structures of these two astrophysical bodies share, indeed, the same pressure-temperature domain and are described by the same H/HE EOS \cite{rop10,fortney11}. Developments of ab initio EOS and high pressure experiments are currently very active fields, being  motivated by the discovery of numerous exoplanets.

\subsubsection{Effect of rotation and magnetic fields}

 Untill recently, eclipsing binaries were thought to provide the best validation test of theoretical models for stellar/substellar internal structure and evolution. If the components of a binary system orbit in a plane which is along the line of site of an observer, a so-called eclipsing binary, it is possible to determine with high accuracy the masses and radii of both components. This provides an observed mass-radius relationship  which can be compared to model predictions and test fundamental physics like the EOS implemented in the structure models. Eclipsing binaries are thus considered as the best astrophysical laboratories to test the interior physics of  
stars and brown dwarfs. Many efforts have been and are still  devoted to the detection and analysis of eclipsing binary systems. While enough observations now exist to test the low mass star regime down to the bottom of the Main Sequence \cite{ribas06}, with many more systems expected from the Kepler mission (see e.g \cite{slawson11}), brown dwarf eclipsing binaries remain extremely rare, with only a few systems known (\cite{stassun06, irwin10}). 

But  the idea strongly  entrenched in the community that eclipsing binaries provide the best tests for EOS and interior structure models based on "standard" physics has recently been called into question.   
This new challenge yields one of the most recent  novelty regarding the physics of low mass stars and brown dwarfs. It highlights the effect of rotation and/or magnetic fields on the inner structure of fully convective objects, effects which are usually not included in "standard" models. Because eclipsing binaries are fast rotators and very active objects, these systems are indeed the best targets to discover and investigate the effect of rotation and magnetic fields on interior structures. 
As the importance of the latter processes was first highlighted with the  analysis of eclipsing binary systems in the low mass star regime,  I will first discuss these findings before turning to brown dwarfs. This  will help discussing, in the second part of this section, the young brown dwarf eclipsing binary system  2M0535-0546 discovered by \cite{stassun06}. This system shows the puzzling property that the more massive component ($\sim 0.055 \msun$) has a cooler effective temperature than the less massive component ($\sim 0.035 \msun$). Such temperature reversal cannot be explained by "standard" models.

The whole story started with increasing evidences for systematic differences between the observed fundamental properties of low-mass stars in eclipsing binaries and those predicted by stellar structure models (see e.g \cite{ribas06, morales10}). 
 
Particularly, radii and effective temperatures computed from models are 5-10\% lower and 3-5\% higher than observed, respectively.  These differences are significant, given the high accuracy of empirical measurements reached nowadays (typically 1-2\%) \cite{morales10}. 
Problems with atmospheric opacities have first been invoked
as the source of the discrepancy. Opacities, however, have a
modest impact on the stellar radius for these compact stars. Changing the metallicity in
the atmosphere by a factor 100 affects the radius by a factor $\sim$ 7\%, so that
the opacity of eclipsing binaries should have to be increased to an unrealistic level to
yield the observed 10\% effect on the radius \cite{chabrier97}. Missing opacities thus seem to be unlikely
to explain the radius discrepancy. 

Because eclipsing binaries are fast rotators
and magnetically very active, 
a possible explanation for the radius discrepancy is the inhibition of internal convection, due to rotation and/or magnetic field, yielding a reduction of the internal heat flux and thus a smaller contraction during evolution \cite{mullan01,chabrier07}. These effects are described in 1D stellar evolution codes by phenomenological approaches, using either the mixing length parameter \cite{chabrier07} or a  magnetic inhibition parameter \cite{mullan01} to control the efficiency of convection.  In addition to internal convection inhibition,  surface spot coverage could contribute to
a reduction of the internal heat flux, also yielding a larger radius \cite{chabrier07}.
These scenarios thus provide an appealing explanation for the larger radius in rapidly rotating, very active stars. 
The value of the equilibrium field inferred in the phenomenological approach of \cite{chabrier07} to hamper convection is consistent with the observationally determined value of magnetic fields of very low mass stars \cite{reiners07} and with the one obtained with 3D resistive magneto-hydrodynamic (MHD) simulations \cite{browning08}. 
Convection becomes more and more efficient and adiabatic with decreasing mass in this regime, because of increasing mean density and average interior opacities. The aforementioned decreasing convective efficiency due to magnetic fields is thus expected to be relatively less and less consequential as one moves along the mass sequence from
the Sun to the bottom of the main sequence. Such a behaviour is indeed supported by
observations (see Fig. 1 in \cite{chabrier07}). 

\begin{figure}[!h]
\sidecaption
\includegraphics[scale=.35]{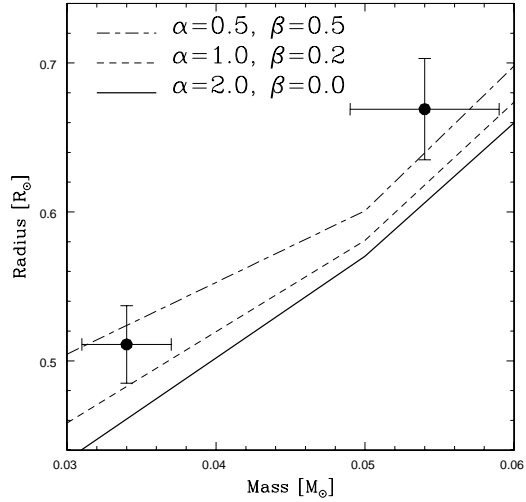}
\caption{Effect of a variation of the mixing length parameter $\alpha$ and the factional surface area $\beta$ covered by cool spots on the mass radius relationships of young brown dwarfs. The black data points are the empirical mass and radius values of the binary brown dwarf 2M0535-0546. Models are from \cite{chabrier07} at an age of 1 Myr, corresponding to the inferred age of 2M0535-0546 \cite{stassun06}.}
\label{binaryBD}       
\end{figure}

These very same effects of magnetically driven inhibition of convection
and spot coverage could also provide a plausible explanation for the temperature
reversal observed in the brown dwarf eclipsing binary system 2M0535-0546 \cite{stassun06}. Interestingly enough, the primary of this system is a faster rotator than the secondary and displays H$_\alpha$ emission 
at a  level 7 times stronger than the emission from the secondary \cite{reiners07b}. This brings support to the aforementioned scenario, with the primary being more affected by magnetic fields, yielding an increase of its radius and decrease of its effective temperature \cite{mohanty09}. To further comfort this idea, \cite{chabrier07} predict a spot coverage of  20\%-30\% to reproduce the fundamental properties of this binary brown dwarfs (see Fig. \ref{binaryBD}). Based on high resolution spectroscopy of the primary, \cite{mohanty10} find that $\sim$ 70\% spot coverage is required for the primary in order to explain the mismatch between the effective temperature inferred form the TiO-$\epsilon$ band and that from KI absorption feature. In comparison, \cite{morales10} find that a spot coverage of $\sim$ 35\% provides an overall good agreement between models and observations for several low mass stars (0.2-0.8 $\msun$) in eclipsing binary systems. While these numbers look consistent among all analysis, adding support to the effect of spot coverage, very recently, \cite{mohanty12} casted doubts on this interpretation. They obtained high resolution spectroscopy for the secondary of the young brown dwarf binary 2M0535-0546 and found the same discrepancy between $\te(TiO-\epsilon)$ and $\te(KI)$ as reported for the primary. If spots are responsible for the $\te$ mismatch found in both the primary and the secondary, they cannot explain the $\te$ reversal of the primary. Consequently, if magnetic fields were still responsible for the latter,  only convection inhibition in the interior of the primary could be the physical mechanism to solve the problem, but not spot coverage. The issue is far from being settled and this  intriguing case deserves more follow-up.

\begin{figure}[!h]
\sidecaption
\includegraphics[scale=.35]{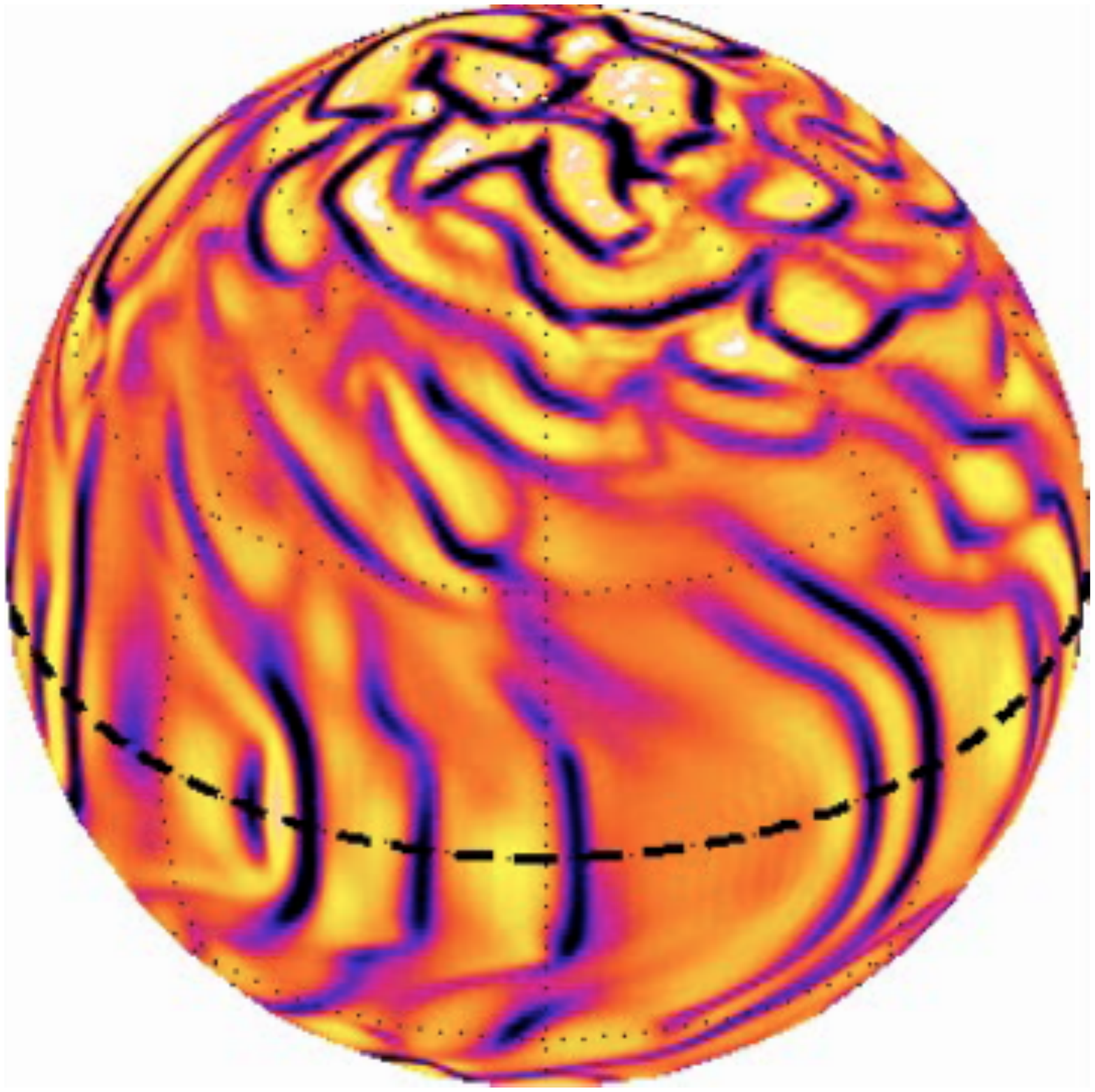}
\caption{Global view of the radial velocity $v_{\rm r}$  on spherical surface deep into the interior of a fast rotating, fully convective M-dwarf or brown dwarf. Upward flows are reddish and downward flows are blue-ish \cite{browning13}. Courtesy M. Browning.}
\label{vbrowning}       
\end{figure}

Multi-dimensional hydrodynamical and magneto-hydrodynamical simulations seem to offer a promising avenue to explore those complex processes \cite{browning08, showman13}. 
The effect of rotation on convective flows is illustrated in Fig. \ref{vbrowning}, showing how rotation organises convection into organised rolls in a fully convective object. The interior rotation profile is constant on cylinders parallel to the rotation axis, reflecting the Taylor-Proudman constraint \cite{browning08, browning13, showman13}. The MHD simulations of \cite{browning08} show that the magnetic field also strongly impacts the convective flows, reducing the development of differential rotation established in pure hydrodynamical simulations. Magnetic fields can thus play a major role in establishing the interior profile of a fully convective object, yielding some weakening of the convection \cite{browning08}. Those results go along the lines of the predictions of \cite{chabrier07}.  More theoretical and numerical works are clearly required to explore these thrilling issues. 

\section{The very early evolution of Brown dwarfs}
\label{accr}

\subsection{Birth and  growth}

The broad picture of star and brown dwarf formation starts with a molecular cloud which collapses and fragments, giving birth to prestellar or pre-brown dwarf cores which in turn collapse and produce central protostars or proto-brown dwarfs surrounded by an envelope and an accretion disk. The details are much more complex and are still intensively studied \cite{mckee07, luhman12}. Several mechanisms have been proposed for the formation of brown dwarfs (see review by \cite{luhman12} and references therein). More and more observational evidences tend to support the idea that brown dwarfs form like stars and arise from the smallest prestellar cores (\cite{luhman12}).  The recent detection of a pre-brown dwarf core of $\sim 0.02-0.03 \msun$ \cite{andre12} and of proto-brown dwarf candidates of a few Jupiter masses in Taurus \cite{palau12}, if confirmed, would add another credit to  the idea that brown dwarf formation is a scaled-down version of low mass star formation.  This is also confirmed by the detection that young brown dwarfs have disks e.g \cite{harvey12} and outflows e.g \cite{joergens12a, joergens12b}, like young stars.

A new paradigm based on the idea of episodic accretion is now superseding the standard picture of steady or
slowly-varying accretion as a function of time onto a protostar or a proto-brown dwarf. 
The standard picture  describes the prestellar core collapse as a quasi-static process, giving rise to a constant accretion rate $c_s^3/G \sim 10^{-6} \rm {M_\odot yr^{-1}} $, with $c_s$ the speed of sound and $G$ the gravitational constant \cite{mckee07}. An important failure of the standard model is the well known "luminosity problem" related to the fact that accretion at such rates produces accretion luminosities ($L_{\rm acc} \propto {G M_\star \dot M \over R_\star}$, where $M_\star$ and $R_\star$ are respectively the mass and luminosity of the central accreting object) factors of 10-100 higher than typically observed  for protostars embedded within a  massive envelope. This classic problem  \cite{kenyon90} was recently aggravated by large surveys of star forming regions based on the Spitzer Space Telescope which revealed the presence of a large population of low luminosity embedded sources. The observed protostar luminosity distributions in all surveyed star forming regions are found  to be strongly inconsistent with the standard model. They are better explained by non-steady accretion processes, with long quiescent phases of accretion interrupted by short episodes of high accretion (see e.g. \cite{evans09} and references therein). 

\begin{figure}[!h]
\sidecaption
\includegraphics[scale=.70]{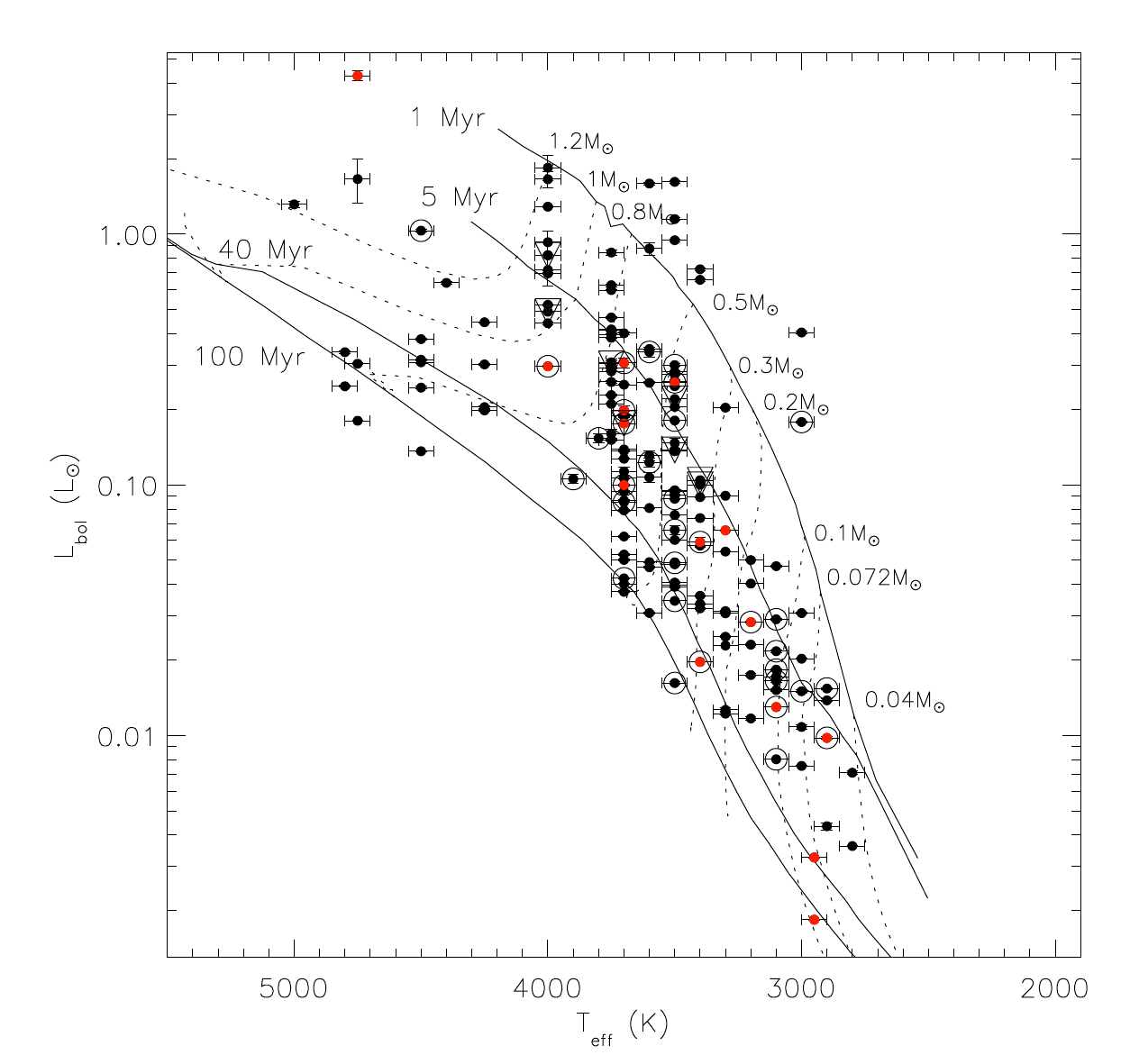}
\caption{Illustration of the observed luminosity spread in the star forming region $\lambda$ Orionis ($\sim$ 5 Myr)  \cite{bayo12}. Courtesy A. Bayo.}
\label{hrbayo}       
\end{figure}

Episodic accretion may not only solve the classic luminosity problem
of embedded sources (with typical ages $<$$<$1 Myr), but also another
intriguing feature observed in young clusters of a few Myr, namely the
luminosity spread observed in their luminosity - effective temperature
diagram (the Hertzsprung-Russell diagram).  The origin of this
luminosity spread, illustrated in Fig. \ref{hrbayo}, is highly controversial.
Whether it arises from observational uncertainties, physical processes
or stems from a real age spread among young objects belonging to the
same cluster, are crucial questions for the understanding of star/brown dwarf
formation. The interpretation of the luminosity spread as an age
spread is used as an argument in favour of slow formation
process, taking several Myr to tens of Myr. This view conflicts with
our current understanding of star formation characterised by much
shorter timescales, typical of shock-dominated turbulence.

\subsection{Effects of accretion on the evolution}
\begin{figure}[!h]
\sidecaption
\includegraphics[scale=.35]{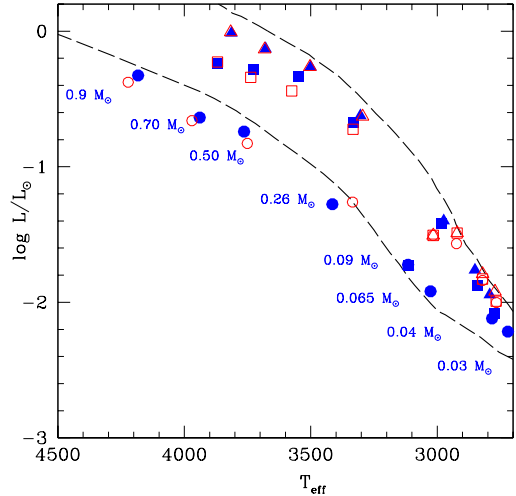}
\caption{Illustration of the luminosity spread predicted by 
theoretical models including effects of episodic accretion. The coloured symbols show the position of objects having the same age (1 Myr) but different accretion histories. 
They correspond to a coeval population of low mass objects. The two long-dashed
 curves indicate 1 Myr and 10 Myr isochrones (from \cite{baraffe12}).}
\label{hraccretion}       
\end{figure}

To explain the luminosity spread above mentioned,  \cite{baraffe09} recently suggested the idea that non
steady accretion at very early stages of evolution, during the
embedded phases, could still strongly impact the structure of young
low mass stars and brown dwarfs even after a few Myr, i.e. after the main phase of accretion. Accretion effects yield significantly more compact structures, i.e. smaller radii compared to those of non accreting objects of same mass and age. This contraction stems from the increase in gravitational energy as mass is added, yielding higher central pressures and temperatures compared to the non accreting case. During high accretion bursts, the accretion timescale $\tmdot=M/\dot M$ remains smaller than the thermal timescale of the accreting object $\tkh = GM^2/(RL)$. Its structure has thus no time to
adjust to the incoming mass and energy, and the radius remains smaller than the non accreting
counterpart of same mass and age. The more compact structure
of the accreting object result in a fainter luminosity compared to the non accreting counterpart. The former  thus looks "older" than the latter, as
illustrated in Fig. \ref{hraccretion}.

Those effects are not only able to produce a luminosity spread, as shown in Fig. \ref{hraccretion}, similar to the observed one but can also
explain other observational features  like unexpected lithium depletion in some young objects \cite{baraffe10} or the
properties (mass and radius) of FU Ori objects characterised by strong bursts of accretion \cite{baraffe12}. 
This scenario can thus provide an explanation to the luminosity spread which is a better concept than the idea of an age spread,  being in better agreement with current observations and with current understanding of star formation. 
An interesting effect on young brown dwarf evolution, due to the combination of accretion effects and deuterium nuclear burning, was noted by  \cite{baraffe12}. Because the energy released from deuterium fusion can partly overcome the gravitational energy increase due to mass accretion for objects with masses $\simle 0.04 \msun$, accretion is predicted to have less effect on the structure of those low mass brown dwarfs. This must remain true above the deuterium burning minimum mass, i.e for $M \simgr 0.01 \msun$. Consequently, within this scenario, the luminosity spread in the mass range $\sim 0.01-0.04 \msun$ is predicted to  be smaller than for masses below and above.  This prediction could be  tested by observations to confirm or not the idea of episodic accretion.

The effects of accretion above described  are
still debated \cite{hosokawa11} as their description relies on several modelling assumptions. This now requires 
more robust physical foundations. The origin of the luminosity spread
is  still an open issue, which should motivate further theoretical
and observational studies, given its impact on the global
understanding of star and brown dwarf formation. In particular, on the observational
front, it is now crucial to study a variety of star forming
environments and the properties of their low mass stars and brown dwarfs combining
alternative age indicators (position in a luminosity- effective temperature diagram, gravity measurement,
rotational properties, properties of accretion disks, etc...) to test
current ideas about  early accretion.

\subsection{Brown dwarf or planet?}

Another topical debate is how to  observationally distinguish a brown dwarf from a planet. 
This question is relevant if characterising these two families of objects by their formation processes rather than using the IAU definition based on the deuterium burning minimum mass. The IAU definition is admittedly practical from an observational point of view but is arbitrary with respect to the formation process. Because the two families of objects share a similar mass domain between a few  $M_{\rm Jup}$ and tens of $M_{\rm Jup}$, defining clear diagnostic enabling to distinguish a genuine brown dwarf from a planet is a key problem in the field. It is however not an an easy task \cite{chabrier07b}. For young objects, the luminosity has been suggested as a possible signature of the formation process.
The idea was floated with the work of \cite{marley07} suggesting that young planets, formed in a protoplanetary disk through the core accretion model, a widely accepted planet formation scenario, should be fainter than predicted by earlier models. This stems from the assumption made in the work of \cite{marley07} that all the energy liberated in the accretion shock produced by matter falling onto the forming planet's surface is radiated away and does not contribute to the planet's energy balance (and thus to its luminosity). On the other hand, brown dwarfs 
 forming like stars via gravitational collapse, as described in the previous sections, should be significantly more luminous. 
 Assuming that a fraction of the accretion energy is absorbed by the proto-brown dwarf and contributes to its intrinsic luminosity, brown dwarfs should
 start from a high initial specific entropy state, the so-called "hot start".

\begin{figure}[!h]
\sidecaption
\includegraphics[scale=.35]{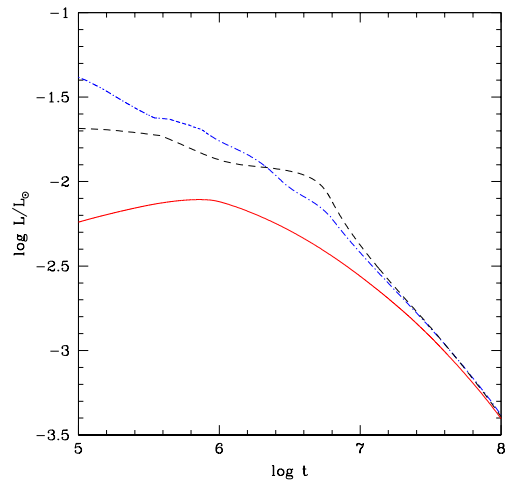}
\caption{Evolution of the luminosity as a function of time of  0.04 $\msol$ brown dwarfs with different accretion histories. The red solid line corresponds to an object formed via "cold" accretion ($\alpha=0$) and the blue dash-dotted curve corresponds to "hot" accretion ($\alpha=0.2)$ (from the models of \cite{baraffe12}). The parameter $\alpha$ corresponds to the fraction of accretion energy absorbed by the accreting object. The black dashed line shows the evolution at constant mass.}
\label{tlm04}       
\end{figure}

 Within this scenario, the measured luminosity of young planetary mass objects (provided their age can be estimated) could thus reveal key information on  their formation process. Measuring luminosities of such low mass objects is now a reality with the new generations of adaptive optics
systems, such as VLT-SPHERE \cite{claudi06} or the Gemini Planet Imager (GPI  \cite{macintosh06}), or with the coming successor of the Hubble Space telescope (the JWST  \cite{gardner06}) and with the perspective of the European Extremely Large Telescope (EELT \cite{gilmozzi07}). The idea is thus exciting since it  might provide an accessible way to disentangle a brown dwarf from a giant planet. 
 Unfortunately, this scenario is  oversimplified
 (see e.g. \cite{mordasini12}). Accretion
 processes onto forming planets and forming brown dwarfs, despite
 operating under very different conditions, may share similar
 properties regarding the fraction of accretion energy radiated away
 and absorbed by the central object. As illustrated in Fig. \ref{tlm04}, different properties of accretion can produce young brown dwarfs with a range of initial luminosities, depending on the accretion rate and the fraction of accretion energy absorbed by the central object. 
 Consequently, the luminosity of young planetary mass objects
 certainly tells something about the physics of the accretion process. Whether this diagnostic  can reveal
 anything about the formation process itself (i.e. gravitational
 collapse versus core accretion) is however debatable and requires more work on the physics of accretion.

\section{ The future of brown dwarf physics}

This brief overview of recent issues in the physics of brown dwarfs shows a lively field with new questions that calls for  further developments and new ideas in atmosphere, dust and internal structure modelling.  The treatment of dust and clouds continues being a major issue, requiring increasing level of sophistication in the modelling. This topic is also becoming of high interest for  the understanding of exoplanet atmospheres with an apparent prevalence of dust in those objects, like in  the hot Jupiter HD 189733b \cite{pont13}. Because of the remarkable level of accuracy of brown dwarf observations, compared to currently available data for exoplanets, one may bet that progress in ultra cool atmosphere modelling will preferentially come from the brown dwarfs community. A new interest is also emerging in the study of variability and "meteorological" processes in brown dwarf atmospheres, problems which are usually restricted to the analysis of planet atmospheres.  Processes like cloud disruption can translate into observable signatures that observers are  tracking to provide additional constraints  on the physical properties of clouds (size, distribution, conditions for destruction, etc..) \cite{buenzli12, radigan12}. The study of and search for lightning can also provide interesting information on the physical properties of dust in cool atmospheres \cite{helling13}.  This field develops rapidly due to improved observational techniques to monitor the variability of very faint objects . 
The exoplanet community can learn a lot from the experience gained on brown dwarfs to understand the atmospheric properties of their favourite objects.   
In parallel to the continuous development in dust/cloud treatment, major progress has been made and is still coming regarding fundamental physics, with improved line lists for major molecular absorbers. At the time of this writing, a new methane line list is   expected from the Exomol project \cite{tennyson12}. It is crucially needed for T/Y dwarfs and exoplanets
and may solve the unexplained behavior of near-infrared colours for the coolest T and Y dwarfs mentioned in Sect.  \ref{cmd}.  Further mandatory improvement in the micro physics concerns  calculation of molecular line broadening under pressures and temperatures  relevant to those cool and dense atmospheres. Current treatment of broadening factors are very rough
 and mostly rely on  models valid at lower pressures and higher temperatures  \cite{homeier05}.  Line broadening, however,   plays an important role in shaping the spectra of brown dwarfs. The possibility now to perform accurate comparison between models and observations should motivate more experimental and theoretical studies  of line
 broadening. 
  
Interior structure and evolutionary models for brown dwarfs have considerably improved within the past decade and have now reached their limit based on 1D stellar evolutionary codes and phenomenological approaches. The field is  entering a new era based on multi-dimensional magneto-hydrodynamical models.  Various tools are available or being developed  in the community, which will provide a better description and understanding of  complex, but ubiquitous, physical processes like convection, rotation and magnetic fields. Among the available  numerical tools, we can quote the anelastic code ASH \cite{browning08}, which  filters out sound waves and linearises thermodynamical fluctuations around a background reference state. Anelastic solvers are consequently restricted to the study of flows with velocities much smaller than the sound speed, i.e with very low Mach numbers. They can be applied to most  brown dwarf interior processes. Other numerical tools  are being developed, based on e.g time implicit methods \cite{viallet11, kifonidis12}, which are  numerically less limited than anelastic methods and can study low to moderate Mach number processes on long time scales, relevant to evolutionary problems. Such tools can study accretion effects on the structure of convective objects in order to better describe the redistribution in the interior of mass and  energy accreted  onto the surface of an object. These processes are currently studied  through phenomenological approaches, assuming    
 instantaneous and uniform redistribution of  mass and internal energy brought by the accreted material. Great advancement have also been made and are still expected on the front of formation theories, with three-dimensional MHD simulations including radiation hydrodynamics being now underway to explore  the second collapse and the physics of accretion at very early stages of evolution \cite{masson12}. All those numerical developments will  provide a consistent picture of early and later stages of evolution of brown dwarfs, promising a wealth of exciting and novel results that should further motivate  investment in the field of brown dwarf physics. 
 
 To conclude, brown dwarfs, as failed stars, are sometimes considered as the "Ugly Duckling" with an image tarnished by the big nuclear-powered stars and the appealing exoplanets. But brown dwarfs are great physical laboratories. Their discovery, as a cosmic confirmation of the effects of matter degeneracy predicted by quantum mechanics,  marks a  page of modern physics history.  Still glowing with youth, they will keep surprising us for a long time.

\end{document}